\documentclass{article}
\usepackage{url}
\usepackage[pdftex]{graphicx}
\usepackage{amsmath}
\usepackage{booktabs}
\usepackage{amsfonts}
\usepackage{float}
\usepackage{tcolorbox}

\usepackage{authblk}  

\title{\textbf{Query, Don't Train:} \\ Privacy-Preserving Tabular Prediction from EHR Data via SQL Queries}
\author[]{Josefa Lia Stoisser\textsuperscript{*}}
\author[]{Marc Boubnovski Martell\textsuperscript{*}}
\author[]{Kaspar Märtens}
\author[]{Lawrence Phillips}
\author[]{Stephen Michael Town}
\author[]{Rory Donovan-Maiye}
\author[]{Julien Fauqueur}
\affil[]{Novo Nordisk}

\date{May 2025}

\begin{document}
\maketitle

\begin{abstract} Electronic health records (EHRs) contain richly structured, longitudinal data essential for predictive modeling, yet stringent privacy regulations (e.g., HIPAA, GDPR) often restrict access to individual-level records. We introduce \textbf{Query, Don’t Train} (QDT): a \textbf{structured-data foundation-model interface} enabling \textbf{tabular inference} via LLM-generated SQL over EHRs. Instead of training on or accessing individual-level examples, QDT uses a large language model (LLM) as a schema-aware query planner to generate privacy-compliant SQL queries from a natural language task description and a test-time input. The model then extracts summary-level population statistics through these SQL queries, and the LLM performs chain-of-thought reasoning over the results to make predictions. This inference-time–only approach enables prediction without supervised model training, ensures interpretability through symbolic, auditable queries, naturally handles missing features without imputation or preprocessing, and effectively manages high-dimensional numerical data to enhance analytical capabilities. We validate QDT on the task of 30-day hospital readmission prediction for Type 2 diabetes patients using a MIMIC-style EHR cohort, achieving F1 = 0.70, which outperforms TabPFN (F1 = 0.68). To our knowledge, this is the first demonstration of LLM-driven, privacy-preserving structured prediction using only schema metadata and aggregate statistics—offering a scalable, interpretable, and regulation-compliant alternative to conventional foundation-model pipelines. \end{abstract}

\noindent\textsuperscript{*} Equal contribution. Listing order is random.

\section{Introduction}
EHRs store richly structured, longitudinal data spanning diagnoses, laboratory results, procedures, medications, and outcomes—resources that are critical for predictive modeling and clinical decision support \cite{kim2019evolving, tsai2025harnessing}. However, regulations such as the U.S. HIPAA Privacy Rule and the EU GDPR impose strict safeguards for protected health information, including consent, minimization, and access controls, with substantial legal and institutional constraints on data use \cite{cohen2018hipaa, voigt2017eu}. These policies often prohibit direct access to patient-level records, creating significant barriers for model development, particularly in cross-institutional settings where data-sharing agreements are difficult to establish or enforce.

Despite these constraints, public datasets such as MIMIC-III have enabled research in EHR-driven prediction under carefully controlled conditions, supporting tasks such as mortality forecasting, hospital readmission risk, and treatment efficacy modeling \cite{johnson2020mimic, meng2022interpretability,zhu2025designing}. Traditional supervised models—especially tree-based methods like XGBoost—continue to dominate tabular prediction tasks due to their robustness to heterogeneous features, irregular target functions, and missing data \cite{grinsztajn2022tree, NEURIPS2024_751ef1e7, mcelfresh2023neural}. Transformer-based in-context learners, such as TabPFN, offer classification via training-set conditioning, though they still require access to raw examples at inference time \cite{hollmann2022tabpfn, den2024context, qu2025tabicl, bai2023transformers}.

LLMs have recently demonstrated strong performance both in structured reasoning tasks, including text-to-SQL translation \cite{gao2023text,stoisser2025sparks}, and tabular prediction tasks \cite{hegselmann2023tabllm, kim2025table}. Recent advancements have introduced privacy-preserving techniques and agent-based frameworks to address these challenges \cite{liu2025advances}.  Deep learning models can be trained across decentralized datasets using federated learning, enabling collaborative prediction without sharing sensitive data \cite{abadi2016deep,chua2024mind, kuang2024federatedscope, wang2025federated}. Agent-based frameworks allow models to autonomously perform multi-step reasoning over structured data, facilitating complex clinical decision-making \cite{liu2025advances}.

LLMs have recently demonstrated strong performance both in structured reasoning tasks, including text-to-SQL translation \cite{gao2023text}, and tabular prediction tasks \cite{hegselmann2023tabllm, kim2025table}. Recent advancements have introduced privacy-preserving techniques and agent-based frameworks to address these challenges \cite{liu2025advances}. Deep learning models can be trained across decentralized datasets using federated learning, enabling collaborative prediction without sharing sensitive data \cite{kuang2024federatedscope, wang2025federated}. Additionally, advancements in neural network-based techniques emphasize algorithmic strategies for learning while safeguarding sensitive information through differential privacy \cite{abadi2016deep,chua2024mind}. Another promising approach is CRYPTEN, a software framework that enables secure multi-party computation (MPC) for machine learning, allowing parties to collaboratively train models on private datasets while ensuring data privacy \cite{knott2021crypten}. Agent-based frameworks allow models to autonomously perform multi-step reasoning over structured data, facilitating complex clinical decision-making \cite{liu2025advances}.

These capabilities suggest a new opportunity: using LLMs not just for text generation, but for \textbf{schema-aware query planning} that operates under privacy constraints. SQL serves as a controlled, interpretable interface that enables LLMs to retrieve relevant aggregate statistics—without exposing individual-level data—thereby preserving compliance with HIPAA and GDPR \cite{cohen2018hipaa, voigt2017eu}.

In this work, we introduce \textbf{Query, Don’t Train}, a two-stage, framework for clinical tabular prediction without direct access to raw EHR data. Our approach is grounded in three pillars:
\begin{itemize}
    \item \textbf{Privacy preservation}, by ensuring only policy-compliant SQL queries are issued and no patient-level data is revealed.  
    \item \textbf{Structured reasoning}, which derives interpretability from two key sources: (1) LLM-mediated chain-of-thought predictions over query results, and (2) the symbolic, auditable queries themselves.
    
    \item \textbf{Robustness to missing data}, as the model dynamically selects and conditions on available features at inference without imputation.  
\end{itemize}

We validate our approach on 30-day readmission prediction in a MIMIC-style cohort for Type 2 diabetes patients, showing that it obtains an F1-score of 0.70 while offering interpretability and compliance out of the box.

\begin{figure*}[ht]
\vskip 0.2in
\begin{center}
\includegraphics[width=\textwidth]{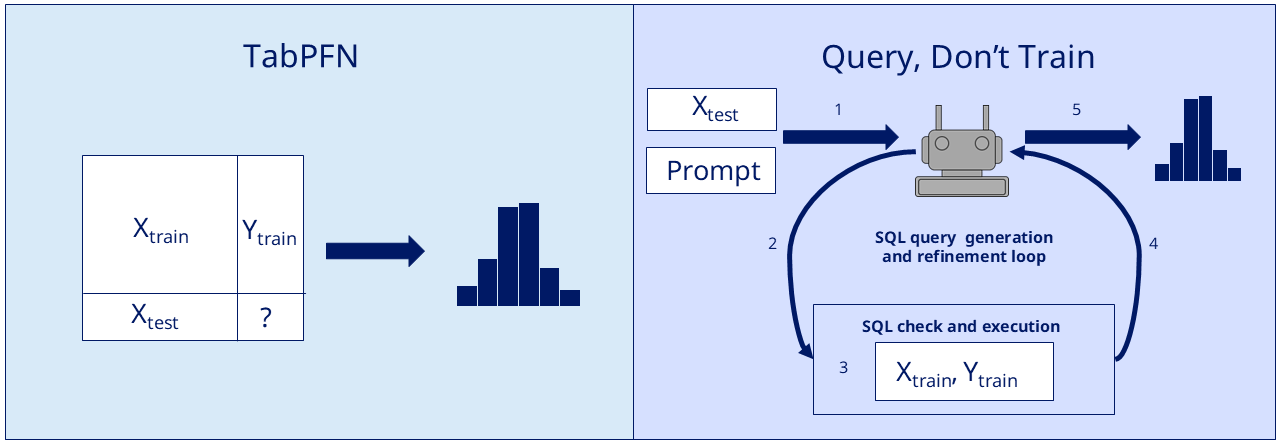}
\caption{\textbf{Comparison of TabPFN and our "Query, Don’t Train" (QDT) approach.} TabPFN uses the training set directly during inference. In contrast, QDT follows: (1) receive test record and task prompt, (2) generate SQL queries, (3) enforce compliance with privacy policies, (4) execute approved queries to retrieve summary statistics, (5) predict using chain-of-thought reasoning. QDT enables privacy-preserving, interpretable inference without raw data access.
}
\label{icml-historical}
\end{center}
\vskip -0.2in
\end{figure*}

\section{Experiments}

\subsection{Experimental Setup} \label{setup}

In our experimental setup, we utilize OpenAI's o4-mini model as the LLM agent, which also serves as the basis for our LLM-only baseline. To implement the agent, we leverage the LangChain library\footnote{\url{https://python.langchain.com/api_reference/community/agent_toolkits/langchain_community.agent_toolkits.sql.toolkit.SQLDatabaseToolkit.html}} . For an example of a run, please refer to Appendix \ref{app:ex}, where we provide details on each SQL command used for interpretability.

To comply with privacy policies, we restricted queries via the system prompt to provide only summary-level statistics, defined as data aggregated and averaged over two or more patients. This constraint is enforced using a separate LLM to validate that only queries requesting summary-level statistics are allowed to proceed to execution. The LLM is prompted with a predefined list of rules (e.g. no queries on cohorts, limited repeated or overlapping queries, monitor query patterns) and can determine whether a query meets the specified criteria. In practice, this validation would be implemented through a firewall to prevent unauthorized data access \cite{kruse2017security}.

\subsection{Datasets}
We focus on predicting 30-day hospital readmissions for patients with Type 2 Diabetes in US hospitals \cite{diabetes_130-us_hospitals_for_years_1999-2008_296}\footnote{\url{https://www.kaggle.com/c/1056lab-diabetes-readmission-prediction/data}}. The dataset consists of patient records $x_i$, which include demographics, laboratory results, procedures, and prior admissions, with binary outcome labels $y_i \in \{0, 1\}$.  We conducted K-fold cross-validation with 5 iterations, utilizing a subset of 2,000 randomly sampled patients in each fold. With approximately 12\% of patients readmitted within 30 days in our evaluation subset, this approach effectively addresses the imbalanced nature of the readmission task.

\subsection{Baselines}
We compare our method against three baselines: TabPFN~\cite{hollmann2022tabpfn} is a pre-trained transformer-based predictor trained to perform tabular classification by conditioning on the training set at inference time. It is particularly relevant as it accesses $\mathcal{D}_{\text{train}}$ during inference, similar in spirit to our method, albeit without privacy constraints. XGBoost~\cite{chen2016xgboost} is a widely-used gradient boosting framework for tabular data. We train XGBoost on the training set $\mathcal{D}_{\text{train}}$ and evaluate it on the test set, representing the standard supervised learning baseline with full access to training data. Additionally, we compare our method with an LLM-only baseline that receives only $x^{\text{test}}$ and a prompt containing the problem formulation.

\subsection{Classification Results}
We compare our approach against TabPFN~\cite{hollmann2022tabpfn} and XGBoost~\cite{chen2016xgboost}. Despite never accessing the raw data, our method achieves competitive performance in predicting 30-day readmissions, as indicated by the metrics presented in Table~\ref{tab:diabetes_readmission}. Specifically, our Query, Don't Train methodology demonstrates strong precision and recall, underscoring the effectiveness of structured reasoning over aggregate statistics. These results highlight the potential of our approach to provide accurate predictions while utilizing minimal training resources.

\begin{table}[t]
\footnotesize
\caption{Performance comparison of different models on 30-day readmission prediction for Type 2 Diabetes patients predicted for a subset of 2,000 patients. Evaluation metrics include Precision, Recall, and F1-score. The errors are represented as standard deviations (±). Query, Don't Train (QDT) refers to using SQL queries to perform predictions without direct access to patient-level data.}
\label{tab:diabetes_readmission}
\centering
\begin{tabular}{lcccccc}
\toprule
\textbf{Model} & \textbf{Precision} & \textbf{Recall} & \textbf{F1-score} \\
\midrule
TabPFN & $0.63 \pm \scriptstyle{0.05}$ & $\textbf{0.76} \pm \scriptstyle{0.07}$ & $0.69 \pm \scriptstyle{0.06}$ \\
XGBoost & $0.65 \pm \scriptstyle{0.04}$ & $0.68 \pm \scriptstyle{0.06}$ & $0.66 \pm \scriptstyle{0.05}$ \\
LLM & $0.54 \pm \scriptstyle{0.03}$ & $0.51 \pm \scriptstyle{0.04}$ & $0.52 \pm \scriptstyle{0.03}$ \\
\midrule
QDT & $\textbf{0.68} \pm \scriptstyle{0.05}$ & $0.73 \pm \scriptstyle{0.06}$ & $\textbf{0.70} \pm \scriptstyle{0.05}$ \\
QDT \newline (30\% less features) & $0.65 \pm \scriptstyle{0.04}$ & $0.69 \pm \scriptstyle{0.05}$ & $0.67 \pm \scriptstyle{0.04}$ \\
QDT \newline (70\% less features) & $0.62 \pm \scriptstyle{0.05}$ & $0.65 \pm \scriptstyle{0.04}$ & $0.64 \pm \scriptstyle{0.04}$ \\
\bottomrule
\end{tabular}
\end{table}

\subsection{Ablation Study on Missing Features}

To investigate the impact of feature availability on model performance, we conducted an ablation study by systematically removing features from $x^{\text{test}}$. The findings illustrate that our method maintains robust performance even with reduced feature sets. When 30\% of the features were omitted, the performance metrics showed only a modest decrease in the F1-score, dropping to 0.67. This demonstrates that, despite missing features, the agent effectively utilized the remaining features in $x^{\text{test}}$ to identify relevant similar examples, which it uses to reason for accurate predictions. When a feature like 'age' is missing, the LLM omits age-based filters and instead generates queries using only available features, enabling robust predictions without imputation. However, with a substantial reduction of 70\% of features, the performance was impacted more significantly, resulting in an F1-score of 0.64. These results attempt to solve the challenges posed by incomplete data in real-world EHR scenarios \cite{NEURIPS2024_751ef1e7}.

\section{Conclusion}

This work introduces QDT, a new framework that reimagines structured prediction through symbolic interaction rather than model training. Our findings demonstrate that LLMs can serve as foundation models for structured data without requiring access to raw examples or parameter tuning. By pairing LLM-generated SQL queries with cohort-level aggregation and chain-of-thought reasoning, QDT constructs implicit, task-conditioned table representations entirely at inference time. This paradigm offers a practical and conceptually distinct alternative to pretraining: it scales across tasks with no model updates, provides interpretability through auditable query outputs, and complies with privacy regulations by design.

The approach is particularly suited to high-stakes domains like healthcare, where individual-level data is sensitive and institutional data-sharing is often infeasible. QDT offers clear advantages in deployment flexibility, explainability, and robustness to missing data, as the system dynamically selects what to query based on feature availability. These attributes make it a compelling candidate for real-world clinical decision support under strict data governance. While demonstrated in healthcare, this abstraction readily extends to other structured domains such as finance, education, and public policy.

In sum, QDT represents a step toward a new class of foundation model interfaces for structured data—ones that emphasize reasoning over memorization, and symbolic querying over supervised optimization.

\section{Limitations and Future Work}

Despite these strengths, several limitations must be addressed. First, the computational efficiency of LLM-driven query generation remains uncertain, particularly as tasks become more complex. Inference time for QDT increases with the number of features, schema complexity, and database size, as each additional element may require extra queries and reasoning. Second, while our experiments focus on structured tabular data, extending this framework to multi-modal EHRs (e.g., imaging or unstructured clinical notes) may require further innovations in prompt engineering and query design.

The privacy constraints we implement allow access only to aggregated results for two or more patients. These constraints can be adjusted to enforce stricter censoring policies, and more fine-grained privacy-preserving mechanisms can be incorporated as needed. Our k-anonymity constraint offers limited privacy and is vulnerable to inference attacks; future work will explore combining this method with Differential Privacy for stronger guarantees. The privacy agent also blocks queries on small cohorts, repeated or overlapping queries, and monitors query patterns to prevent differencing attacks and other indirect disclosures.

Another consideration is the potential for adversarial or suboptimal queries generated by LLMs, which pose risks to the reliability of QDT, particularly in healthcare. To enhance reliability, we are developing automated query validation and error detection to mitigate these risks. Ensuring the reliability of the query-generation process in high-stakes environments is crucial for future work. Additionally, although our method has been tested in US hospitals, broader validation across diverse healthcare systems is essential for establishing generalizability. A few-shot LLM baseline with anonymized in-context examples would provide a fairer comparison, and we plan to include this in future evaluations.

\bibliographystyle{plain} 
\bibliography{bib} 

\appendix

\section{Example Agent Run} \label{app:ex}
This section provides an example agent run. The example demonstrates how the LLM produces interpretable outputs. The interpretability comes from the reasoning being shown step-by-step, which helps users understand how the model arrives at its conclusion. Moreover, SQL queries are generated and executed as part of this process, allowing users to trace back the logic and validate intermediate reasoning steps. Finally, the uncertainty of the model’s predictions is visible in the final step, where the SQL query results reflect the confidence in the outcome. This ensures that the outputs are explainable and not just black-box results.

\begin{tcolorbox}[colback=yellow!5!white, colframe=yellow!75!black, title=Step 1: Prompt]
\textbf{Task:} Use a given schema and patient record to predict if the patient is likely to be readmitted. This involves querying the database using SQL for patients with similar feature values and inferring the outcome.  

\vspace{5pt}

\textbf{Input Sample:}  
\begin{verbatim}
Patient Record: 
  ...
  time_in_hospital: 8
  num_lab_procedures: 71
  num_procedures: 6
  num_medications: 37
  max_glu_serum: 1.0
  ...
\end{verbatim}

\vspace{5pt}

\textbf{Database Schema:} 
\begin{verbatim}
Table: diabetes
Columns: 
  ... time_in_hospital, num_lab_procedures, num_procedures, 
  num_medications, max_glu_serum, readmitted ...
\end{verbatim}
\end{tcolorbox}

\begin{tcolorbox}[colback=gray!5!white, colframe=gray!75!black, title=Step 2: Agent Thought]
I have the schema of the diabetes table and the patient record features. The key features of the patient record seem to be \texttt{time\_in\_hospital}, \texttt{num\_lab\_procedures}, \texttt{num\_procedures}, \texttt{num\_medications}, \texttt{max\_glu\_serum}. To predict if the patient will likely be readmitted, I can find similar patient records in the database with similar feature values and check their readmission status. I will query the diabetes table for records with same values for these key features and infer the likely outcome based on the \texttt{readmitted} column.  
\end{tcolorbox}

\begin{tcolorbox}[colback=green!5!white, colframe=green!50!black, title=Step 3: Action]

\vspace{5pt}

\begin{verbatim}
SELECT time_in_hospital, num_lab_procedures, num_procedures, 
       num_medications, max_glu_serum, readmitted  
FROM diabetes  
WHERE time_in_hospital = 8.0  
  AND num_lab_procedures = 71.0  
  AND num_procedures = 6.0  
  AND num_medications = 37.0  
  AND max_glu_serum = 1.0  
LIMIT 10  
\end{verbatim}

\end{tcolorbox}

\begin{tcolorbox}[colback=blue!5!white, colframe=blue!75!black, title=Step 4: Output]

\begin{verbatim}
[]
\end{verbatim}

\end{tcolorbox}

\begin{tcolorbox}[colback=gray!5!white, colframe=gray!75!black, title=Step 2: Agent Thought]

\vspace{5pt}

\textbf{Thought:} The query to find exact matches for the patient’s key features returned no results. This means there are no exact matches for all these values in the database. I should relax the query to find patients with similar values. For example, patients with \texttt{time\_in\_hospital} close to 8, \texttt{num\_lab\_procedures} close to 71, \texttt{num\_procedures} close to 6, \texttt{num\_medications} close to 37, and \texttt{max\_glu\_serum=1}. Then I can check their readmission status to infer the likely outcome for this patient.  

\vspace{5pt}

\end{tcolorbox}

\begin{tcolorbox}[colback=green!5!white, colframe=green!50!black, title=Step 3: Action]

\begin{verbatim}
SELECT time_in_hospital, num_lab_procedures, num_procedures, 
       num_medications, max_glu_serum, readmitted  
FROM diabetes  
WHERE time_in_hospital BETWEEN 6 AND 10  
  AND num_lab_procedures BETWEEN 60 AND 80  
  AND num_procedures BETWEEN 4 AND 8  
  AND num_medications BETWEEN 30 AND 40  
  AND max_glu_serum = 1.0  
LIMIT 10  
\end{verbatim}

\end{tcolorbox}

\begin{tcolorbox}[colback=blue!5!white, colframe=blue!75!black, title=Step 4: Output]

\begin{verbatim}
{data_samples}
\end{verbatim}

\end{tcolorbox}

\begin{tcolorbox}[colback=gray!5!white, colframe=gray!75!black, title=Step 2: Thought]

Now let's check the readmission status of these patients.  

\end{tcolorbox}

\begin{tcolorbox}[colback=green!5!white, colframe=green!50!black, title=Step 3: Action]

\begin{verbatim}
SELECT readmitted, COUNT(*) as count  
FROM diabetes  
WHERE time_in_hospital BETWEEN 6 AND 10  
  AND num_lab_procedures BETWEEN 60 AND 80  
  AND num_procedures BETWEEN 4 AND 8  
  AND num_medications BETWEEN 30 AND 40  
  AND max_glu_serum = 1.0  
GROUP BY readmitted  
\end{verbatim}

\end{tcolorbox}

\begin{tcolorbox}[colback=blue!5!white, colframe=blue!75!black, title=Step 4: Output]

\begin{verbatim}
[(0,160),(1,140)]
\end{verbatim}

\end{tcolorbox}

\begin{tcolorbox}[colback=gray!5!white, colframe=gray!75!black, title=Step 5: Thought]
Among patients with similar features to the given patient (\texttt{time\_in\_hospital} 6--10, \texttt{num\_lab\_procedures} 60--80, \texttt{num\_procedures} 4--8, \texttt{num\_medications} 30--40, \texttt{max\_glu\_serum=1}), 160 were not readmitted and 140 were readmitted. The majority were not readmitted, so the patient is more likely not to be readmitted.  

\vspace{5pt}

\textbf{Final Answer: No  }

\end{tcolorbox}


\section{Hyperparamters XGBoost} \label{app:hyper}

In this study, we utilized the \textbf{XGBoost} library \cite{chen2016xgboost} in Python to perform 30-day readmission prediction for Type 2 Diabetes patients. To enhance the model's performance, we conducted systematic hyperparameter tuning utilizing the \texttt{GridSearchCV} class from the \textbf{Scikit-learn} library \cite{pedregosa2011scikit}. 

The following hyperparameters were tuned, along with their respective ranges:

\begin{itemize}
    \item \textbf{Learning Rate (eta)}: Explored in the range of 0.01 to 0.3. The learning rate controls the contribution of each new tree to the overall prediction, where a smaller value generally requires more boosting rounds and allows the model to learn more cautiously, reducing the risk of overshooting optimal parameter values.
    
    \item \textbf{Max Depth}: Tested values ranged from 3 to 10. This parameter affects the complexity of the individual trees, with deeper trees capable of capturing intricate patterns at the potential cost of increased overfitting.
    
    \item \textbf{Min Child Weight}: Values were varied from 1 to 10. This parameter sets a minimum threshold for the sum of instance weights required in a child node, thus helping to control overfitting.
    
    \item \textbf{Subsample}: Evaluated rates of 0.5, 0.7, and 1.0. This parameter determines the fraction of the training data used to grow each tree, with lower values potentially diminishing overfitting through randomization.

    \item \textbf{Colsample\_bytree}: Investigated values included 0.3, 0.5, and 0.8, indicating the fraction of features that are sampled for each individual tree.
\end{itemize}

The hyperparameter tuning process involved performing a grid search combined with \textbf{5-fold cross-validation}, which ensured a thorough assessment of model performance across various hyperparameter combinations. The optimal configuration identified through this process was subsequently used to train the final model, facilitating improved predictive accuracy in the 30-day readmission outcomes among patients with Type 2 Diabetes.

\end{document}